# A Multi-Criteria Framework with Voxel-Dependent Parameters for Radiotherapy Treatment Plan Optimization

Masoud Zarepisheh, Andres F. Uribe-Sanchez, Nan Li, Xun Jia, and Steve B Jiang

Center for Advanced Radiotherapy Technologies and Department of Radiation Medicine and Applied Sciences, University of California San Diego, La Jolla, CA 92037-0843, USA

E-mails: sbjiang@ucsd.edu, xunjia@ucsd.edu.

**Abstract**

In a treatment plan optimization problem for radiotherapy, a clinically acceptable plan is usually generated by an optimization process with weighting factors or reference doses adjusted for organs. Recent discoveries indicate that adjusting parameters associated with each voxel may lead to better plan quality. However, it is still unclear regarding the mathematical reasons behind it. To answer questions related to this problem, we establish in this work a new mathematical framework equipped with two theorems. The new framework clarifies the different consequences of adjusting organ-dependent and voxel-dependent parameters for the treatment plan optimization of radiation therapy, as well as the different effects of adjusting weighting factors versus reference doses in the optimization process. The main discoveries are threefold: 1) While in the organ-based model the selection of the objective function has an impact on the quality of the optimized plans, this is no longer an issue for the voxel-based model since the entire Pareto surface could be generated regardless the specific form of the objective function as long as it satisfies certain mathematical conditions; 2) A larger Pareto surface is explored by adjusting voxel-dependent parameters than by adjusting organ-dependent parameters, possibly allowing for the generation of plans with better trade-offs among different clinical objectives; 3) Adjusting voxel weighting factors is preferred to adjusting the voxel reference doses since the Pareto optimality can be maintained.

## 1. Introduction

Treatment planning in cancer radiation therapy can be treated as a decision-making problem. Planner seeks for a treatment plan to deliver a certain amount of prescription dose to a cancerous target, while sparing dose to nearby critical structures and organs at risks. The apparent conflicts between those objectives make a multi-criteria decision-making technique an appropriate tool to solve this problem. The main idea of the multi-criteria technique is to reduce the size of candidate set of plans by considering only those plans, called Pareto plans, for which it is impossible to improve some objectives without worsening others. The next step is to look for a clinically acceptable plan under the trade-offs between different criteria among the set of Pareto plans, referred to as Pareto surface. There exist several methods to handle this complicated procedure, such as fine-tuning optimization parameters in a trial-and-error fashion or by some heuristic approaches (Xing *et al.*, 1999a; Xing *et al.*, 1999b), pre-computing a well discrete representative of Pareto surface and navigating among them (Küfer *et al.*, 2000; Cotrutz *et al.*, 2001; Hamacher and Küfer, 2002; Lahanas *et al.*, 2003; Küfer *et al.*, 2003; Craft *et al.*, 2006; Thieke *et al.*, 2007; Shao and Ehrgott, 2008; Craft and Bortfeld, 2008; Monz *et al.*, 2008; Hong *et al.*, 2008; Craft and Monz, 2010; Craft *et al.*, 2012), and prioritizing the evaluation criteria to avoid sacrificing those goals of more importance while improving those less important ones (Langer *et al.*, 2003; Jee *et al.*, 2007; Wilkens *et al.*, 2007; Deasy *et al.*, 2007; Clark *et al.*, 2008; Long *et al.*, 2012; Falkinger *et al.*, 2012). Among them, fine-tuning optimization parameters is arguably the most common paradigm.

In the optimization parameter-tuning regime, a commonly used approach is to adjust organ-dependent parameters, such as the weighting factors and the prescription doses for the PTV and the thresholds for the critical organs. The planner tries to achieve clinical goals, like dose-volume constraints, by manipulating the organ-dependent parameters. Many biological-based and dose-based evaluation criteria have also been introduced by researchers into the radiotherapy optimization, e.g. minimum and maximize dose, mean dose, equivalent uniform dose (EUD), generalized equivalent uniform dose (gEUD), tumor control probability (TCP), and normal tissue complication probability (NTCP) (See Romeijn *et al.*, 2004 and references therein). Clinical experiments have demonstrated a great impact of using different evaluation criteria in the optimization process on the optimized plan quality (Craft *et al.*, 2005; Kessler *et al.*, 2005; Xia *et al.*, 2005). From the mathematical viewpoint, different criteria result in different Pareto surfaces. Romeijn *et al.* (2004) and Hoffmann *et al.* (2008) showed that some organ criteria generate the same Pareto surface; however, there are still a wide variety of criteria leading to different Pareto surfaces. So far there is no general agreement on the choice of the objective function and it still remains unanswered as for the question of "which evaluation criterion leads to a better treatment plan?" In this paper, we will show that this issue can be withdrawn by exploiting voxel-based optimization model. In our model, a large Pareto surface exist and different parts of this surface are explored by applying different set of organ evaluation criteria. Moreover, the entire Pareto surface is explored by a voxel-based optimization model, as long as the voxel penalty terms and their derivatives are increasing functions.

The second motivation of this work is from the recent research works about voxel-based optimization models (Cotrutz and Xing, 2002, 2003; Wu *et al.*, 2003; Yang and Xing, 2004; Breedveld *et al.*, 2007; Shou *et al.*, 2005), in which one tunes parameters associated to each voxel in the objective function, as opposed to the organ-based model in which all voxels within a specific organ are tied together and treated equally in the objective function. It has been demonstrated that the voxel-based optimization tends to result in more qualified plans compared to the conventional organ-based approaches for clinical cases. Moreover, voxel-based optimization planning also facilitates interactive treatment planning by providing specific access to certain parts of the dose-volume histogram (DVH) curves and isodose layouts of interest. If a specific part of the DVH or isodose curves does not comply with the clinician's requirements, the planner can pick out the involved voxels and adapt their contributions in the objective function (Cotrutz and Xing, 2002, 2003). Up to now, there are two different schemes for adjusting voxel penalty terms in the objective function: voxel weighting factor adjustment (Cotrutz and Xing, 2002, 2003; Wu *et al.*, 2003; Yang and Xing, 2004; Breedveld *et al.*, 2007; Shou *et al.*, 2005) and voxel reference dose[1] adjustment (Lougovski *et al.*, 2010; Wu *et al.*, 2003).

Despite the success achieved by the voxel-based model in clinical studies, those models are mainly proposed heuristically and the fundamental reasons regarding the efficacy of them is unclear. In this paper we aim at building a mathematical framework for this voxel-based optimization approach, which enables us to answer the following questions naturally raised regarding this model: 1) Why do voxel-based models lead to better plan quality than organ-based models? 2) How much improvement in DVH curves can be expected by utilizing the voxel-based approach compared to the conventional organ-based approach? In particular, can we expect to improve some parts of the DVH curves without worsening other parts? 3) What is the appropriate voxel-based objective function? 4) What is the difference, in terms of plan quality, between adjusting voxel weighting factors and voxel reference doses, and which one is preferred?

This paper unfolds as follows. The new framework is presented in Section 2. Section 3 elaborates on the results of the new framework and clarifies the differences between the variants of parameter adjustment. Finally, Section 4 is devoted to the conclusions and future researches.

**2. A New Framework for Treatment Optimization**

*2.1 Three Different Pareto Surfaces*

Pareto optimality is an inevitable part of the inverse treatment planning when approached from the direction of multi-criteria optimization. It helps us to get rid of the so-called non-Pareto plans that are not worth consideration. Generally, the Pareto optimal solutions are those feasible ones for which it is impossible to improve some criteria without deteriorating others. Equivalently speaking, improvement at no cost is possible for non-Pareto solutions, while this is impossible for Pareto ones.

---

[1] It has been referred to as the prescription dose in literature.

Pareto optimality is defined with respect to specific criteria used in evaluating the plan quality. In intensity modulated radiation therapy (IMRT) literature, Pareto optimality has been conventionally defined based on the evaluation criteria (objective functions) associated to the organs. At least one evaluation criterion is given to each organ and a plan is called Pareto optimal if there does not exist another plan that is better in terms of at least one criterion and not worse with respect to every other criterion. For example, if we consider the maximum dose as an evaluation criterion for each organ at risk (OAR) and the minimum dose for the planning target volume (PTV), then a plan is Pareto optimal if it is impossible to decrease the maximum dose in one OAR without increasing the maximum dose in at least one other OAR or decreasing the minimum dose in PTV.

Since the traditional definition of Pareto optimality in IMRT is based on the given evaluation criteria, the set of the Pareto solutions (Pareto surface) would depend on the specific evaluation criteria used in the studies, and it is not clear in advance which Pareto surface includes a plan with clinically better trade-offs. For example, a TCP/NTCP-based Pareto surface may contain a better plan for one patient, while a gEUD-based Pareto surface may include a better plan for another one.

Here, we define Pareto optimality using the DVH and dose distribution as the evaluation criteria are the most common standards to evaluate the plan quality. For the sake of simplicity, we treat the PTV over-dosing and under-dosing equally; however, the results could be generalized easily.

1- ($X_{OEC}$): A treatment plan is called *organ evaluation criteria Pareto* (OEC Pareto), if improvement in some organ evaluation criteria is only possible at the cost of another organ evaluation criterion. For instance, the organ evaluation criteria could be the maximum dose to OAR and the minimum dose to the PTV. In this case, the set of all Pareto treatment plans (Pareto surface) is denoted by $X_{OEC}$.

2- ($X_{DD}$): A treatment plan is called *dose distribution Pareto,* if it is impossible to improve the radiation doses in some voxels without worsening those in other voxels (improvement of doses in PTV voxels means getting closer to the prescribed dose, while that in voxels in OARs means delivering less radiation). Let $X_{DD}$ denote the set of all Pareto treatment plans in this case.

3- ($X_{DVH}$): A treatment plan is called *DVH Pareto,* if we cannot improve a certain part of the DVH curve of an organ without deteriorating either other part of that DVH or the DVH curves of other organs. Similarly, we denote the set of all Pareto plans in this case by $X_{DVH}$.

In fact, the OEC Pareto surface is the commonly used one in current IMRT treatment planning. In the next two subsections we will show that how we can explore these surfaces and also we will prove their relationship. To facilitate reading, hereafter we refer to dose distribution Pareto optimality as Pareto optimality, and dose distribution Pareto surface as Pareto surface.

*2.2 Exploring Pareto Surfaces*

In commonly used organ-based treatment planning, at least one evaluation criterion (e.g., maximum, mean, or minimum organ dose) is associated with each organ, and a desired plan is obtained by finding an appropriate tradeoff between these criteria. All voxels within a specific organ are weighted equally in each organ criterion. Mathematically speaking, each criterion is a function of the corresponding voxel doses in this organ that is invariant under permutation of voxel indices. As opposed to the organ-based model, a voxel-based model allocates non-uniform penalty to the voxels. Let us consider a typical IMRT inverse planning problem where the fluence map $x$ is the decision variable. Here we just consider IMRT optimization problem for the sake of simplicity, and the proposed framework can be easily applied to any other treatment plan optimization problems (e.g., volumetric modulated arc therapy).

Problems (1) and (2) demonstrate a typical voxel-based and an organ-based model, respectively:

$$x(w) = \arg\min_{x \geq 0} \sum_{\sigma \in S} \sum_{j \in v_\sigma} w_j F_j (D_j x), \tag{1}$$

$$\bar{x}(w) = \arg\min_{x \geq 0} \sum_{\sigma \in S} w^\sigma G^\sigma (D^\sigma x), \tag{2}$$

where $S = T \cup C$ is the set of structures with $T$ accounting for the tumor and $C$ for critical structures, $v_\sigma$ denotes the set of voxels belong to the structure $\sigma$. $w_j$ and $w^\sigma$ are the weights corresponding to the voxel $j$ in the voxel-based model and the structure $\sigma$ in the organ-based model. $D$ denotes the dose deposition matrix and its entry $D_{jk}$ specifies the dose received by the voxel $j$ from a beamlet $k$ at its unit intensity. $D_j$ is the $j$ th row of matrix $D$ that corresponds to voxel $j$, and $D^\sigma$ is the set of rows corresponding to the organ $\sigma$. $F_j$ is a voxel penalty function, and $G^\sigma$ is an organ penalty function.

In Problem (1), each voxel has its own penalty contribution in the objective function with its specific parameter such as $w_j$ defining its importance. For target voxels, it is preferred to have dose close to the prescription value $r^\sigma$, while low radiation dose is desirable for the voxels belong to OARs. Therefore, the penalty functions for voxels in OARs should be the increasing functions of dose, and penalty functions for tumor's voxels should be increasing functions of deviation to the prescribed value.

The following theorem reveals that we can explore the OEC and dose distribution Pareto surfaces by employing organ-based and voxel-based models respectively. The proof of this theorem is given in the Appendix.

***Theorem 1:***

a- The optimal solutions of Problem (2) are OEC Pareto with respect to the criteria $G$; i.e.,
$$\bigcup_{w>0} \bar{x}(w) \subset X_{OEC}$$

b- Let $F_j$ be increasing functions for each $\sigma \in C, j \in v_s$, and increasing functions of $|D_j^\sigma x - r^\sigma|$ for each $\sigma \in T, j \in v_s$, then
$$\bigcup_{w>0} x(w) \subset X_{DD}$$

c- Let $F_j$ and its derivative be increasing functions for each $\sigma \in C, j \in v_s$, and increasing functions of $|D_j^\sigma x - r^\sigma|$ for each $\sigma \in T, j \in v_s$, and the derivative of $F$ be positive on its domain, then
$$\bigcup_{w>0} x(w) = X_{DD}$$

The first part of the above theorem shows that some parts of the OEC Pareto surface can be explored by organ-based model. In fact, we are able to generate almost all Pareto points which are called *properly* Pareto solutions and we might miss the others referred to as *non-properly* Pareto solutions (Ehrgott, 2005; Miettinen, 1999). The second part of the theorem reveals that if the voxel penalty functions are appropriate increasing functions, then the whole Pareto surface, except for non-properly Pareto points, could be explored by the voxel-based model. The third part of the theorem states that the whole Pareto surface could be explored provided that the voxel penalty functions and their derivatives are appropriate increasing functions and the derivatives are positive on their domains. Moreover, it is easy to show that having the increasing derivatives results in convexity and assures the global optimality.

While choosing the set of appropriate evaluation criteria is a controversial issue for organ-based model, the above theorem tells us this is not anymore an issue for voxel-based model. In fact, almost entire Pareto surface could be explored, as long as the voxel penalty functions are appropriate increasing functions, and the whole Pareto surface could be generated if the derivatives are increasing and positive as well. In particular, this means generating the $X_{DD}$ is irrespective of the specific forms of the penalty functions. For instance, by employing the following popular power penalty function, we are able to explore almost entire Pareto surface by changing the voxel-based weights regardless the value of the exponent $q_\sigma$ providing that $q_\sigma > 1$.

$$\min_{x \geq 0} \sum_{\sigma \in C} \sum_{j \in v_\sigma} w_j (D_j x)^{q_\sigma} + \sum_{\sigma \in T} \sum_{j \in v_\sigma} w_j |D_j x - r^\sigma|^{q_\sigma} \quad (3)$$

Since the derivative of the power function is not positive everywhere in (3), we might miss non-properly Pareto points. In order to catch the entire Pareto surface we can use, for example, the following exponential penalty function.

$$\min_{x \geq 0} \sum_{\sigma \in C} \sum_{j \in v_\sigma} w_j \exp(D_j x) + \sum_{\sigma \in T} \sum_{j \in v_\sigma} w_j \exp(|D_j x - r^\sigma|)$$

*2.3 Relationship between Pareto Surfaces*

The following theorem clarifies the relationship between three different Pareto surfaces that have been introduced in section 2.1. The proof of this theorem is also given in the Appendix.

***Theorem 2:***

If $G^\sigma$ is an increasing function for each $\sigma \in C$, and is an increasing function of $|D^\sigma x - r^\sigma|$ for each $\sigma \in T$, then

$$X_{OEC} \subset X_{DVH} \subset X_{DD} \tag{4}$$

Relation (4) encompasses two important messages. The first message comes from the relation $X_{OEC} \subset X_{DD}$. It implies that a larger Pareto surface is explored by adjusting voxel-based weighting factors than by adapting organ-based weighting factors, possibly leading to a plan with better trade-offs among different clinical criteria. In fact, we are exploring different parts of the large Pareto surface $X_{DD}$ by using different set of the organ evaluation criteria in the organ-based model. That is, the organ-based problem picks certain plans out of the Pareto surface $X_{DD}$ based on the given OEC, and leaves out others as the non-Pareto plans. However, the part of the $X_{DD}$ that is not navigated by this approach may contain some more desirable plans, because there is generally no scientific or fundamental theory behind choosing OEC.

The second message can be drawn from the relation $X_{OEC} \subset X_{DVH}$, which states that the plans generated by the organ-based model are already DVH Pareto, where a uniform weighting factor is assigned to all voxels within an organ. Therefore, it is impossible to improve some parts of the DVH curves without worsening others by allocating the non-uniform weights to voxels. In particular, by adjusting the voxel-based weights after the organ-based weight adjustment, it is impossible to improve treatment plans in terms of DVH criteria at no cost. However, this statement does not exclude the possibility of improving the plan at a small cost (Cotrutz and Xing, 2002, 2003; Wu *et al.*, 2003; Yang and Xing, 2004; Shou *et al.*, 2005).

**3. Treatment Plan Optimization by Using Quadratic Penalty Functions**

In this section, we will elaborate on the results of the above two theorems by considering a specific problem with a popular quadratic penalty function. The comparison between voxel weighting factor adjustment and voxel reference dose adjustment is provided as well. The quadratic dose function has been a very popular choice in IMRT treatment plan optimization. Using a quadratic function makes it possible to get rid of the absolute function in (3), and we can also take advantage of the existing efficient algorithms developed for the quadratic optimization problems (Breedveld *et al.*, 2006). The rest of this section is devoted to the organ-based and voxel-based parameter adjustment by employing a quadratic penalty function.

*3.1 Organ-Dependent Parameter Adjustment*

The following problem illustrates a typical organ-based quadratic optimization model for IMRT:

$$\min_{x \geq 0} \sum_{\sigma \in C} w^\sigma \sum_{j \in v_\sigma} (D_j x - r^\sigma)_+^2 + \sum_{\sigma \in T} w^\sigma \sum_{j \in v_\sigma} (D_j x - r^\sigma)^2 \quad (5)$$

Where $a_+$ denotes the vector $a$ with all negative elements replaced by zeros. The weights $w^\sigma$ (for $\sigma \in S$) are the parameters that we need to fine-tune to get a plan with desirable trade-offs. This can be done by a trial-and-error based approach or by using a heuristic update scheme, e.g. developed by Xing et al. (Xing *et al.*, 1999b; Xing *et al.*, 1999a). For tumor voxels, reference doses ($r^\sigma$, $\sigma \in T$) are the prescribed doses given by the clinicians, but for OARs they are parameters, which are adjusted to reshape the DVH curves. In fact, the objective function focuses on the particular tail parts of the DVH curves for OARs by considering penalty only for doses which are higher than the reference doses.

For an OAR, the penalty function in Problem (5) is not an increasing function, because it does not differentiate the doses that are lower than the reference dose. Hence, the conditions of Theorem 1 are not satisfied and Pareto optimality might be lost. The conditions are satisfied only if the reference doses are equal to zero for all OAR voxels. Problem (6) overcomes this issue by penalizing the doses lower than the reference dose with a linear penalty function and doses higher than the reference dose with an extra quadratic dose function. The conditions of Theorem 1 are then fulfilled for this problem, and hence, the Pareto optimality is guaranteed. In fact, the main idea for reshaping the DVH curves is to focus more on certain parts of the DVH by considering the penalty only for over-doses[2]. Instead of penalizing only for over-doses, Problem (6) reshapes the DVH by penalizing over-doses more than the under-doses.

$$\min_{x \geq 0} \sum_{\sigma \in C} w^\sigma \sum_{j \in v_\sigma} (D_j x + (D_j x - r^sigma)_+^2) + \sum_{\sigma \in T} w^\sigma \sum_{j \in v_\sigma} (D_j x - r^\sigma)^2 \quad (6)$$

---

[2] Doses those are higher than the reference doses.

*3.2 Voxel-Dependent Weight Adjustment*

Cotrutz and Xing (2002) and (2003), and subsequently Yang and Xing (2004) and Shou *et al.* (2005), extended their previous work on organ-based quadratic objective function to the voxel-based quadratic model. Wu *et al.* (2003) proposed their systematic approach to adjust the weighting factors per voxel to modify the initial plan and guide it to clinically acceptable one. Breedveld *et al.* (2007) introduced their voxel-based automatic treatment planning in order to generate a plan that complies with all or at least the most important given DVH and maximum-dose constraints. They all employed the following voxel-based quadratic model:

$$\min_{x \geq 0} \sum_{\sigma \in C} \sum_{j \in v_\sigma} w_j^\sigma (D_j x)^2 + \sum_{\sigma \in T} \sum_{j \in v_\sigma} w_j^\sigma (D_j x - r^\sigma)^2 \quad (7)$$

Since the conditions of Theorems 1 and 2 are satisfied in above problem, we can take advantage of these theorem's results. According to Theorem 1, all plans obtained by weighting factor adoption are Pareto, and moreover, almost the entire Pareto surface could be characterized by changing the weighting factors. Regarding Theorem 2, if the plan is achieved by adjusting the organ-based weights, we cannot expect to improve some parts of the DVH at no cost by employing the voxel-based model; however, we can expect to improve some parts at a reasonable cost since we are able to explore some previously missed parts of the Pareto surface. These facts will be demonstrated in the following case study.

*3.3 Organ-based Model Versus Voxel-based Model: A Case Study*

We studied a gynecologic cancer case that has been planned previously using the Eclipse treatment planning system (Varian Medical System, Palo Alto, CA). The approved plan is replicated by using our in-house GPU-based dose and optimization engines to compare the organ-based and voxel-based models on a fair ground. The organ and voxel weighting factors are adjusted automatically and iteratively based on the deviation from the approved plan (see Li *et al.* (2012) for details about weight adjustments). The results are illustrated in Figure 1, where the solid lines represent DVHs for the plan generated by a voxel-based model, and the dashed lines are for the plan generated by an organ-based model. Comparing these two plans, it is clear that by utilizing non-uniform weights to different voxels in an organ, the sigmoid and rectum doses are improved significantly, and the PTV over-dosing and bladder dose are improved too at a reasonable cost of doses to the body, bowel and pelvic bone marrow (PBM). While it is a clinical question as for which of the two plans is more preferable, voxel weighting factors offers a big freedom to handle the tradeoffs among organs. This occurrence can also be observed in the case studies provided in other voxel-dependent research works (Cotrutz and Xing, 2002, 2003; Wu et al., 2003; Yang and Xing, 2004; Shou et al., 2005)

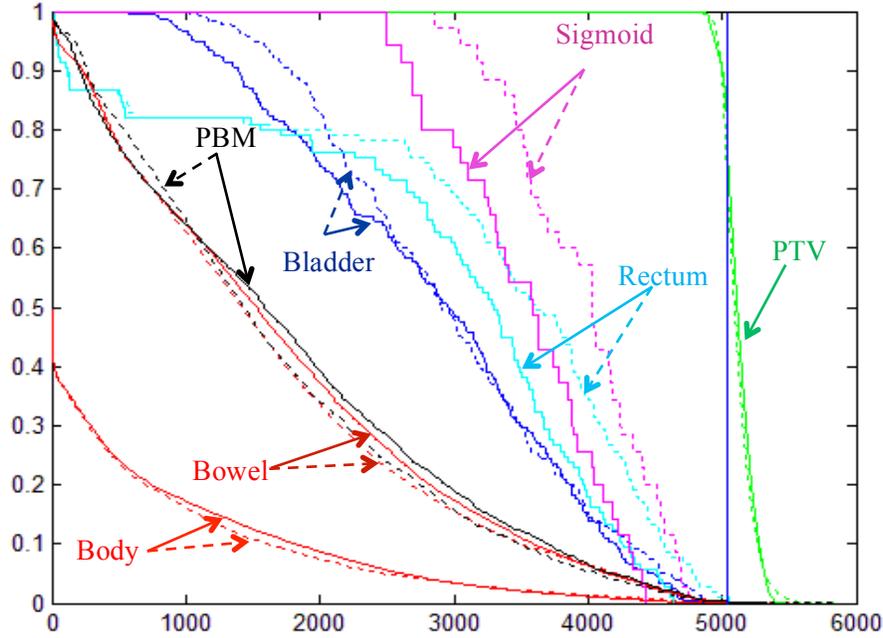

Figure 1: Comparison of the treatment plans optimized with a voxel-based model and an organ-based model for a GYN case. Solid line: voxel-based model; Dashed line: organ-based model. The vertical blue line indicates the prescription PTV dose.

*3.4 Voxel-Dependent Reference Dose Adjustment*

In addition to changing voxel weighting factors, voxel reference doses can also be adjusted as optimization parameters to produce a desirable plan as in the following model:

$$\min_{x \geq 0} \sum_{\sigma \in S} w^{\sigma} \sum_{j \in v_{\sigma}} (D_j x - r_j)^2 \qquad (8)$$

As opposed to the weighting factor adjustment, the Pareto optimality is not guaranteed by adjusting reference doses. In fact, the penalty functions for the OAR and PTV voxels in the above problem are not necessarily the increasing functions of $D_j x$ and $|D_j x - r^{\sigma}|$ which makes the conditions of Theorem 1 unsatisfied.

Apart from the Pareto optimality, Problems (7) and (8) differ in the way that they are adjusting voxel penalty terms in the objective function in order to catch the desirable trade-offs. By fine-tuning voxel-based weighting factors, we are adjusting the quadratic penalty term of each voxel in the objective function, while the linear penalty term of each voxel is adopted by updating the reference doses. To clarify this, let us extend the penalty term for a specific voxel $j$ in Problem (8) as follows:

$$w^{\sigma}((D_j x)^2 - 2r_j(D_j x) + r_j^2)$$

Since $w^\sigma r_j^2$ is independent of the decision variable $x$, it does not affect the optimal solution. Thus, for each voxel there are a quadratic and a linear penalty term in the objective function and only is the linear part altered by updating the reference doses. In contrast, for the cases where weighting factors are adjusted, the quadratic terms are affected. What we can conclude from the aforementioned fact is that the smaller change is expected in the current plan by altering reference doses than weighting factors.

**4. Discussion and Conclusions**

In this paper, we presented two theorems behind the organ- and the voxel-based optimization models by introducing a new mathematical framework for treatment plan optimization. The new framework clarifies the main pros and cons of the different parameter adjustment strategies, and makes it possible to answer some interesting and important questions arisen in these contexts.

The results of the new framework could be summarized as follows:
1- While selecting an appropriate objective function has an impact on the quality of the optimized plan using an organ-based model due to the fact the different objective functions may correspond to different parts of the Pareto surface, the entire Pareto surface could be characterized by a voxel-based model as long as the voxel penalty functions and their derivatives are appropriate increasing functions.
2- Comparing the organ-based and the voxel-based model, the former only captures some parts of the Pareto surface depending on the given organ evaluation criteria. Therefore, we are more likely to get a plan with more desirable trade-offs by exploring the entire Pareto surface via the voxel-based model.
3- By allocating the non-uniform importance factors, rather than the uniform weights, we cannot expect to improve some parts of the DVH at no cost; however, possibly we would be able to improve some parts of the DVH at a small cost by exploring the missing parts of the Pareto surface.
4- In contrary to the voxel weight adjustment, the Pareto optimality is not guaranteed by tuning voxel reference doses. Moreover, since changing the reference doses is equivalent to updating the linear penalty term of each voxel, it is expected that smaller changes will occur in the plan comparing with changing the weighting factors.
5- In organ-based model, when we are changing the reference doses in order to reshape the DVH curves, it is better to consider the linear penalty term rather than the zero penalty for doses lower than the reference dose to assure the Pareto optimality.

Yet, one practical difficulty in the voxel-based optimization models is the dramatically increased number of parameters. We developed a heuristic method to iteratively adjust the voxel weighting factors based on the deviation to the desired DVH. A research topic that needs to be pursued in future is developing an efficient and systematic approach to adjust the weighting factors in order to efficiently navigate through the Pareto surface to choose a clinically acceptable plan.

5   **Acknowledgement**

This work is supported by the University of California Lab Fees Research Program and a Master Research Agreement from Varian Medical Systems, Inc. We would like to thank Dr. Edwin Romeijn for reading this paper and for providing constructive comments.

**Appendix**

Throughout the proofs we need some fundamental results from multi-criteria optimization which can be found in both (Ehrgott, 2005) and (Miettinen, 1999).

*A.1. Proof of Theorem 1:*

*a*) For an arbitrary positive weight vector, an optimal solution of Problem (2) is Pareto with respect to the objective function $G$. It implies that there is no possibility to decrease some $G^\sigma$ without increasing others, which means that the optimal solution is Pareto in terms of the evaluation criteria. Moreover, each properly Pareto point can be generated with some positive weights.

*b*) Let $\bar{x}$ be an optimal solution of Problem (1) for an arbitrary positive weight vector. Then, $\bar{x}$ is a Pareto solution of $\min_{x \geq 0} F(Dx)$. Since function $F_j$ is increasing for each $\sigma \in C, j \in v_s$, and increasing of $|D_j x - r^\sigma|$ for each $\sigma \in T, j \in v_s$, then $\bar{x}$ is a Pareto solution of Problem (A1) as well. Now, by the definition of dose distribution Pareto optimality, it is obvious that each Pareto solution of Problem (A1) is Pareto in terms of the dose distribution, i.e., $\bar{x} \in X_{DD}$.

$$\min_{x \geq 0} \{\{D_j x\}_{\sigma \in C, j \in v_\sigma}, \{|D_j x - r^\sigma|\}_{\sigma \in T, j \in v_\sigma}\} \quad \textbf{(A1)}$$

*c*) Let $\bar{x} \in X_{DD}$ be an arbitrary Pareto solution in terms of the dose distribution. We need to prove that $\bar{x}$ is an optimal solution of Problem (1) for some positive weights. According to the definition of dose distribution Pareto optimality, $\bar{x}$ is a Pareto solution of Problem (A1) which can be converted to the following equivalent linear multi-criteria optimization problem by change of variables (Murty, 1983):

$$\min\{\{D_j x\}_{\sigma \in C, j \in v_\sigma}, \{z_j^+ + z_j^-\}_{\sigma \in T, j \in v_\sigma}\}$$
s.t.
$$D_j x - r^\sigma = z_j^+ - z_j^-, \quad \sigma \in T, j \in v_\sigma \quad \textbf{(A2)}$$
$$x, z^+, z^- \geq 0$$

If we define $z_j^+(\bar{x}) = \max\{0, D_j \bar{x} - r^\sigma\}$, $z_j^-(\bar{x}) = \min\{0, D_j \bar{x} - r^\sigma\}$, then it can be readily shown that $(\bar{x}, z^+(\bar{x}), z^-(\bar{x}))$ is a Pareto solution of Problem (A2). Due to the linearity of (A2), there is a positive weight vector $\bar{w}$ for which $(\bar{x}, z^+(\bar{x}), z^-(\bar{x}))$ solves the corresponding weighted version of the above problem. Now, it is not difficult to show that $\bar{x}$ solves the weighted version of Problem (A1) corresponding to $\bar{w}$. It implies that $\bar{x}$ is a so-called *properly* Pareto solution of this problem. Since $F$ and its derivative are increasing functions with respect to the objective functions of Problem (A1) and the derivative of $F$ is positive on its domain, $\bar{x}$ is a properly Pareto solution of $\min_{x \geq 0} F(Dx)$ as well (Zarepisheh, 2011). So, there are some positive weights for which $\bar{x}$ is an optimal solution of Problem (1) due to the strict convexity of $F$.

*A.2. Proof of Theorem 2:*

First of all, we need to provide a mathematical definition for improvement in DVH curves. Let $\pi(.)$ denote the permutation that sorts every given vector in an ascending order. Then, for a specific structure $\sigma \in C / \sigma \in T$ improvement in DVH curve is equivalent to the componentwise decrease in vector $\pi(D^\sigma x) / \pi(|D^\sigma x - r^\sigma|)$. Now, we prove the relations $X_{OEC} \subset X_{DVH}$ and $X_{DVH} \subset X_{DD}$ separately.

($X_{OEC} \subset X_{DVH}$): If we prove that for each structure $\sigma$, the corresponding objective function $G^\sigma$ decreases as the DVH curve of that structure improves, and $G^\sigma$ does not change as the DVH curve remains unchanged, then the relation $X_{OEC} \subset X_{DVH}$ can be proved easily by contradiction. We prove this property for two different cases $\sigma \in C$ and $\sigma \in T$ individually.

($\sigma \in C$): Let $\bar{x}, \hat{x} \geq 0$ be two feasible solutions of (2). At first, consider the case for which these two plans produce the same DVH curves for structure $\sigma$. For this case we need to show that $G^\sigma(D^\sigma \bar{x}) = G^\sigma(D^\sigma \hat{x})$. The relation $\pi(D^\sigma \bar{x}) = \pi(D^\sigma \hat{x})$, and subsequently $G^\sigma(\pi(D^\sigma \bar{x})) = G^\sigma(\pi(D^\sigma \hat{x}))$, can be deduced from the fact that these two plans generate the same DVH curves for structure $\sigma$. Since the objective function $G^\sigma$ is symmetric and indifferent to the permutation, we have $G^\sigma(D^\sigma \bar{x}) = G^\sigma(\pi(D^\sigma \bar{x}))$, $G^\sigma(D^\sigma \hat{x}) = G^\sigma(\pi(D^\sigma \hat{x}))$, and hence, $G^\sigma(D^\sigma \bar{x}) = G^\sigma(D^\sigma \hat{x})$. Now, consider the case for which plan $\bar{x}$ generates the better DVH curve than plan $\hat{x}$ for structure $\sigma$. In this case, we need to prove that $G^\sigma(D^\sigma \bar{x}) < G^\sigma(D^\sigma \hat{x})$. Since $\bar{x}$ generates the better DVH curve than $\hat{x}$, we have $\pi(D^\sigma \bar{x}) \grave{o} \pi(D^\sigma \hat{x})$ [3]. Moreover, the objective function $G^\sigma$ is an increasing function concluding that $G^\sigma(\pi(D^\sigma \bar{x})) < G^\sigma(\pi(D^\sigma \hat{x}))$. Now, the relation $G^\sigma(D^\sigma \bar{x}) < G^\sigma(D^\sigma \hat{x})$ can be obtained by taking into account the symmetric property of $G^\sigma$.

($\sigma \in T$): The main idea to prove this part is same as the previous part. If $\bar{x}$ and $\hat{x}$ generate the same DVH curves for structure $\sigma$, then $\pi(|D^\sigma \bar{x} - r^\sigma|) = \pi(|D^\sigma \hat{x} - r^\sigma|)$ and so $G^\sigma(\pi(|D^\sigma \bar{x} - r^\sigma|)) = G^\sigma(\pi(|D^\sigma \hat{x} - r^\sigma|))$. Since the objective function $G^\sigma$ is symmetric with respect to the $|D^\sigma x - r^\sigma|$, we have $G^\sigma(|D^\sigma \bar{x} - r^\sigma|) = G^\sigma(|D^\sigma \hat{x} - r^\sigma|)$. Now, if the DVH curve of plan $\bar{x}$ for structure $\sigma$ is better than of plan $\hat{x}$, then

---

[3] $a \grave{o} b \Leftrightarrow a \leq b$ and $a \neq b$

$\pi(|D^\sigma \bar{x} - r^\sigma|)$ ò $\pi(|D^\sigma \hat{x} - r^\sigma|)$. The relation $G^\sigma(|D^\sigma \bar{x} - r^\sigma|) < G^\sigma(|D^\sigma \hat{x} - r^\sigma|)$ can be achieved by taking into account the increasing and symmetric properties of $G^\sigma$ with respect to $|D^\sigma x - r^\sigma|$.

($X_{DVH} \subset X_{DD}$): We argue by contradiction to prove this part. Suppose that $\hat{x}$ does not belong to $X_{DD}$, then we need to prove that $\hat{x}$ does not either belong to $X_{DVH}$. $\hat{x} \notin X_{DD}$ means that there exists another plan like $\bar{x}$ with better amount of radiation in some voxels. Therefore, for each $\sigma \in C / \sigma \in T$ we have $D^\sigma \bar{x} \leq D^\sigma \hat{x} / |D^\sigma \bar{x} - r^\sigma| \leq |D^\sigma \hat{x} - r^\sigma|$ and relation ò holds at least for one structure. Relations $D^\sigma \bar{x} \leq D^\sigma \hat{x}$ and $|D^\sigma \bar{x} - r^\sigma| \leq |D^\sigma \hat{x} - r^\sigma|$ result in $\pi(D^\sigma \bar{x}) \leq \pi(D^\sigma \hat{x})$ and $\pi(|D^\sigma \bar{x} - r^\sigma|) \leq \pi(|D^\sigma \hat{x} - r^\sigma|)$ respectively. It shows that some DVH curves of plan $\bar{x}$ could be improved without deteriorating the others, meaning $\bar{x} \notin X_{DVH}$, which completes the proof.